\documentclass[lettersize,journal]{IEEEtran}
\usepackage{amsmath,amsfonts,amssymb}
\usepackage{algorithmic}
\usepackage{array}
\usepackage{textcomp}
\usepackage{stfloats}
\usepackage{url}
\usepackage{verbatim}
\usepackage{graphicx}
\usepackage{cite}
\usepackage{xcolor}
\usepackage{multirow}
\def\BibTeX{{\rm B\kern-.05em{\sc i\kern-.025em b}\kern-.08em
    T\kern-.1667em\lower.7ex\hbox{E}\kern-.125emX}}
\usepackage{balance}
\usepackage{booktabs}

\begin{document}

\title{CSI-CLIP++: A Scalable Channel Foundation Model for Wireless Communication via CIR-CSI Consistency}

\author{Jun Jiang, Wenjun Yu, Yunfan Li, Yuan Gao, and Shugong Xu \textit{Fellow, IEEE}


\thanks{Jun Jiang and Shugong Xu are with Xi’an Jiaotong-Liverpool University, Suzhou, China, email: jun.jiang25@student.xjtlu.edu.cn, shugong.xu@xjtlu.edu.cn.}

\thanks{Wenjun Yu, Yunfan Li and Yuan Gao are with the School of Communication and Information Engineering, Shanghai University, China, email: yuwenjun@shu.edu.cn, lyf2023@shu.edu.cn, and gaoyuansie@shu.edu.cn.}

\thanks{An earlier version of this paper was presented at the IEEE ICMLCN25\cite{jiang2025mimo}.}

}

\maketitle

\begin{abstract}
Self-supervised learning can exploit large-scale unlabeled channel data to improve the transferability of wireless AI models. Existing channel foundation models are often built on single-domain representations or reconstruction-oriented objectives, which may not explicitly capture the physical correspondence between frequency- and delay-domain channel views. This paper proposes CSI-CLIP++, a scalable channel foundation model for MIMO wireless channels. CSI-CLIP++ treats frequency-domain channel state information (CSI) and delay-domain channel impulse response (CIR) as paired views of the same propagation process and learns transferable representations through CSI-CIR contrastive alignment. The pretrained CSI encoder is adapted to channel identification, beam prediction, and positioning, representing PHY, RAN, and ISAC applications. Experiments on large-scale DeepMIMO scenarios show consistent gains over supervised baselines across environments, carrier frequencies, and data scales. CSI-CLIP++ improves beam prediction Top-1 accuracy by up to 19.31 percentage points and achieves competitive positioning performance, including cross-simulator transfer on a Sionna RT dataset. Backbone scaling results further show that the proposed objective remains effective across encoder architectures and benefits from larger model capacity.

\end{abstract}


\section{Introduction}
\IEEEPARstart{W}{ith} the rapid evolution of 5G-Advanced and the emergence of next-generation wireless networks, wireless communication systems are expected to support increasingly stringent requirements in terms of efficiency, reliability, intelligence, and generalization. Artificial intelligence (AI) has become an important enabling technique for wireless communications and has been widely investigated in tasks such as channel extrapolation\cite{gao2026ai}, channel estimation\cite{li2025survey}, and positioning\cite{pan2025ai}. In particular, the AI-native vision of future wireless networks emphasizes the deep integration of AI into communication systems, where intelligent models are expected to adapt to diverse scenarios and support multiple wireless tasks with limited task-specific supervision. The 3rd Generation Partnership Project (3GPP) has also recognized the potential of AI/ML in the NR air interface, highlighting representative use cases such as CSI feedback enhancement, beam management, and positioning in TR 38.843~\cite{3gpp38843}.

Channel state information (CSI) serves as a fundamental representation of wireless propagation characteristics and plays a central role in many communication and sensing tasks. Typical CSI-related tasks include channel identification, beam prediction, and positioning, which correspond to important applications in physical-layer (PHY) processing, radio access networks (RAN), and integrated sensing and communication (ISAC) systems. Although these tasks have different objectives, they are all closely related to the underlying wireless propagation environment. Therefore, learning general and transferable channel representations is crucial for improving the adaptability of AI models across different wireless scenarios.

Existing AI-based methods for wireless channel processing still face several limitations. Task-specific supervised models usually require large amounts of labeled data and often suffer from performance degradation when deployed in unseen environments, frequency bands, or channel conditions. Multi-task learning methods improve parameter sharing across related tasks, but their performance usually depends on carefully designed task relationships and architectures. Recently, channel foundation models (CFMs) have provided a promising solution by learning general-purpose representations from large-scale wireless data and adapting them to downstream tasks~\cite{jiang2025towards,xu20266g}. However, many existing CFMs mainly focus on single-domain channel representations or reconstruction-based pretraining objectives. Such methods may not fully exploit the physical correspondence between different channel representations, especially the relationship between frequency-domain CSI and delay-domain channel impulse response (CIR).

In wireless channels, CSI and CIR describe the same propagation process from two complementary perspectives. CSI characterizes the channel response in the frequency domain, while CIR reflects the multipath structure in the delay domain. Since these two representations are physically connected through Fourier transformation, they naturally provide paired views of the same channel. Inspired by contrastive representation learning~\cite{radford2021clip}, this property motivates a contrastive pretraining strategy that aligns CSI and CIR in the representation space, allowing the model to learn cross-domain consistent and task-agnostic channel features without relying on task-specific labels or masked reconstruction objectives.

Based on this motivation, this paper proposes CSI-CLIP++, a scalable channel foundation model for multiple-input multiple-output (MIMO) wireless channels. CSI-CLIP++ adopts CSI-CIR contrastive alignment as its self-supervised pretraining objective. By aligning frequency-domain and delay-domain channel representations, the proposed model captures physically consistent channel features and enhances transferability across downstream tasks. Unlike conventional reconstruction-based pretraining methods, CSI-CLIP++ directly learns representation-level correspondence between CSI and CIR, which helps avoid over-reliance on local interpolation-like reconstruction and improves the generality of learned channel embeddings.

In summary, the main contributions of this paper can be summarized as follows:
\begin{enumerate}
    \item We extend our previous conference version, CSI-CLIP, and propose CSI-CLIP++, a scalable CFM for MIMO wireless channels. Compared with the preliminary version, CSI-CLIP++ replaces the ResNet50-based backbone with a Transformer-based ViT-B encoder and further extends the model scale to ViT-L, showing that the proposed contrastive pretraining framework can benefit from larger backbone architectures.

    \item We design a CSI-CIR contrastive alignment objective that exploits the physical correspondence between frequency-domain CSI and delay-domain CIR. By pulling matched CSI-CIR pairs closer and pushing unmatched pairs apart in the representation space, the proposed objective enables cross-domain consistent channel representation learning from large-scale unlabeled CSI-CIR pairs without task-specific labels.
    
    \item We adapt the pretrained CSI encoder to three representative downstream applications, including channel identification, beam prediction, and positioning, corresponding to PHY, RAN, and ISAC applications, respectively. These tasks cover both classification and regression problems and verify the effectiveness of the learned representations across different wireless tasks, prediction types, scenarios, carrier frequencies, and data scales.

    \item We conduct a more comprehensive evaluation than previous version by fixing 7 representative testing scenarios and comparing CSI-CLIP++ with supervised baselines, task-specific models, and existing CFM. Experimental results show that CSI-CLIP++ consistently achieves superior performance, while the ViT-L backbone further improves downstream performance and confirms the scalability of the proposed framework.

\end{enumerate}

The rest of this paper is organized as follows. Section~II reviews related works on AI-driven wireless communication, self-supervised representation learning, and channel foundation models. Section~III introduces the background of the downstream tasks considered in this paper. Section~IV presents the proposed CSI-CLIP++ framework, including the model architecture, CSI-CIR contrastive pretraining strategy, and downstream adaptation process. Section~V describes the experimental setup and evaluates the proposed model on multiple wireless tasks. Finally, Section~VI concludes this paper and discusses future research directions.

\section{Related Works}

\begin{figure}[tp]
\centering\includegraphics[width=0.8\columnwidth]{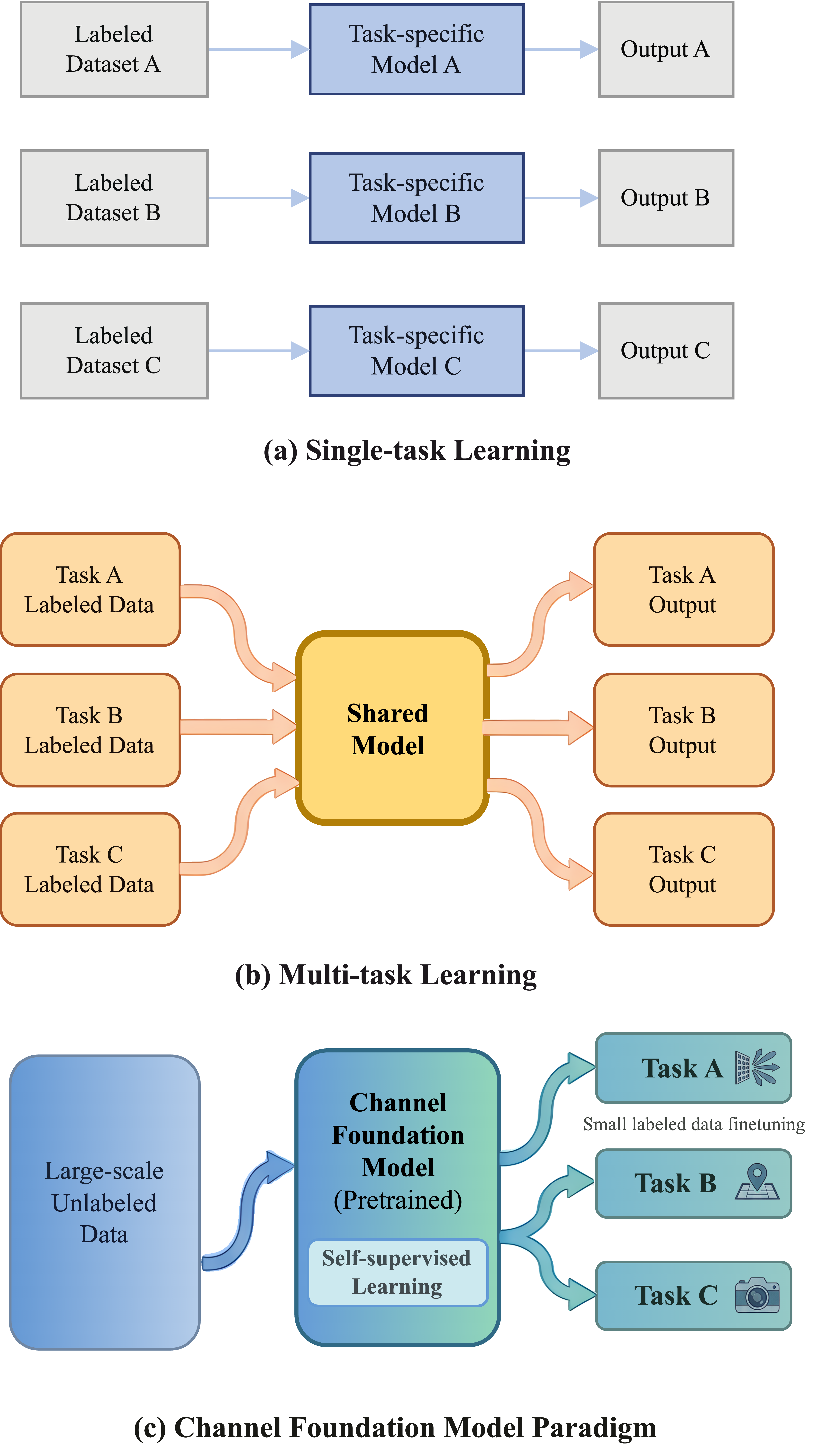}
\caption{\textbf{Evolution of AI-driven wireless communication paradigms:}
(a) Single-task learning relies on isolated task-specific models trained with labeled data. (b) Multi-task learning introduces shared architectures to jointly model multiple related tasks. (c) Channel Foundation Models leverage large-scale self-supervised pretraining and lightweight finetuning to enable generalized representation learning.}
\label{fig:stage}
\end{figure}

\subsection{AI-Driven Wireless Communication Development}
AI-driven wireless communication has evolved from task-specific supervised learning to multi-task learning and, more recently, foundation-model-based paradigms that pretrain reusable representations before downstream adaptation, as illustrated in Fig.~\ref{fig:stage}. Early supervised methods\cite{xu2025enhanced,shi2025channelmamba,wen2018deep} typically follow a ``one task, one model'' design, where separate neural networks are trained for channel estimation, beam prediction, localization, or other channel-related tasks. Although these methods can achieve good performance under fixed experimental settings, they usually depend heavily on labeled data and may generalize poorly to unseen propagation environments, frequency bands, or antenna configurations.

Multi-task learning attempts\cite{jiang2025mtca,wang2023super,jagannath2021multi} to improve data efficiency and parameter sharing by jointly training related wireless tasks with shared model components. By exploiting common structures across tasks, such methods can reduce the need to train independent models for each application. However, their performance often relies on strong task correlations and carefully designed network architectures. When the task relationships are weak or the deployment scenarios become more diverse, multi-task models may still suffer from limited scalability and insufficient generalization.

Foundation models~\cite{zhu2026fm,pan2025large,guler2026multi,jiao2025addressing} provide a more flexible paradigm for wireless intelligence by first learning general-purpose representations from large-scale data and then adapting them to downstream tasks~\cite{cheng2025foundation}. In wireless communications, CFMs aim to capture task-agnostic channel features from CSI, spectrograms, or other radio measurements. Compared with task-specific and multi-task models, this paradigm has the potential to reduce label dependence and improve transferability across different scenarios. Nevertheless, how to design physically meaningful pretraining objectives for structured wireless channels remains an open problem.

\subsection{Self-Supervised Learning for Channel Representation}

Self-supervised learning (SSL) has become an important technique for learning transferable representations from large-scale unlabeled data\cite{gui2024survey}. This paradigm is particularly attractive for wireless communication because massive channel-related data can be collected through simulations, measurements, or network operations, while task-specific labels are often expensive or difficult to obtain. For example, labels for positioning, beam prediction, channel identification, or sensing usually require additional measurement procedures, environmental annotation, or task-dependent processing. Therefore, SSL provides a promising way to exploit unlabeled wireless data and learn general channel representations before downstream adaptation.

Existing SSL methods for wireless representation learning can be broadly divided into reconstruction-based, discrimination-based, and predictive representation learning paradigms. Reconstruction-based methods\cite{xie2022simmim,wang2023droppos} learn representations by recovering missing, corrupted, or masked signal components, and are commonly implemented through masked modeling objectives. Discrimination-based methods\cite{he2020momentum,simclr}, represented by contrastive learning, optimize the representation space by pulling positive pairs closer and pushing negative pairs apart. Predictive representation learning, such as joint-embedding predictive architecture (JEPA)\cite{assran2023self}, predicts the latent representation of masked or unobserved regions rather than reconstructing the raw input. These paradigms provide different ways to exploit the intrinsic structure of wireless signals.

\subsubsection{Masked Channel Modeling}

Masked channel modeling (MCM) is a representative reconstruction-based SSL method for wireless channel pretraining. Inspired by masked language modeling in natural language processing\cite{devlin2019bert} and masked image modeling in computer vision\cite{he2022masked}, it masks part of the observation and trains the model to reconstruct the missing components from the visible regions\cite{hondru2025masked}. Depending on the signal representation, masking can be applied along antenna, subcarrier, time, spatial, or delay dimensions. Because it does not require manual labels and can naturally exploit local correlations in wireless channels, masked modeling has been widely adopted in existing pretraining studies\cite{raviv2026wimamba,sheng2025wireless,catak2025bert4mimo}.

Representative MCM-based CFMs include LWM~\cite{lwm2024} and CSI-MAE~\cite{jiang2026csi}. LWM pretrains a wireless foundation model through masked channel modeling and transfers the learned representation to multiple downstream wireless tasks, while CSI-MAE adopts an autoencoder-style objective that reconstructs masked CSI patches from visible observations. These methods provide important reconstruction-based baselines for CSI representation learning because they exploit unlabeled channel data before downstream adaptation. However, since their pretraining objectives are defined within a single channel representation, they do not explicitly enforce the physical consistency between CSI and CIR.

A key limitation of MCM methods is that they may be less suitable when channel data contain strong local structures or sparse patterns. In MIMO systems, CSI often exhibits structured frequency-antenna correlations, while CIR is usually sparse in the delay domain due to the limited number of dominant propagation paths. Under random masking, the model may recover missing values through local interpolation-like reconstruction rather than learning high-level propagation features. Therefore, a low reconstruction loss does not necessarily imply strong transferability for downstream wireless tasks. This limitation motivates the exploration of representation-level objectives that focus less on raw signal recovery and more on physically meaningful channel consistency.

\subsubsection{Contrastive Learning}
Contrastive learning is another important SSL paradigm for representation learning\cite{hu2024comprehensive}. Unlike MCM, which reconstructs missing signal components, contrastive learning directly shapes the representation space. It encourages representations of positive pairs to be close while pushing apart negative pairs, enabling the model to learn discriminative and invariant features. In wireless communication system, positive pairs can be constructed from different augmentations of the same channel observation, such as noise perturbation, subcarrier masking, antenna-domain cropping, or time-domain distortion. Negative pairs are usually sampled from different users, locations, or propagation environments.

For wireless channel, contrastive learning is particularly suitable when two representations describe the same physical propagation process. Frequency-domain CSI and delay-domain CIR provide such a natural pair. CSI characterizes the frequency response across subcarriers, while CIR reveals the multipath structure in the delay domain. Since they are connected through Fourier transformation, they provide two complementary views of the same channel. Aligning CSI and CIR in the representation space can encourage the model to capture cross-domain consistent propagation features, rather than relying only on single-domain reconstruction. This property makes contrastive alignment especially relevant for learning transferable channel representations in wireless systems.

Despite its potential, contrastive learning also faces challenges in wireless applications. Its performance depends on the construction of positive and negative pairs. If positive pairs are too similar, the pretraining task may become trivial; if they are too different, the model may be forced to align unrelated samples. False negatives may also occur when two samples from nearby locations or similar propagation environments are treated as negative pairs. Therefore, effective contrastive pretraining for wireless channels should be designed based on physical channel relationships rather than purely random data augmentations.

\begin{figure}[t]
\centering\includegraphics[width=1.05\columnwidth]{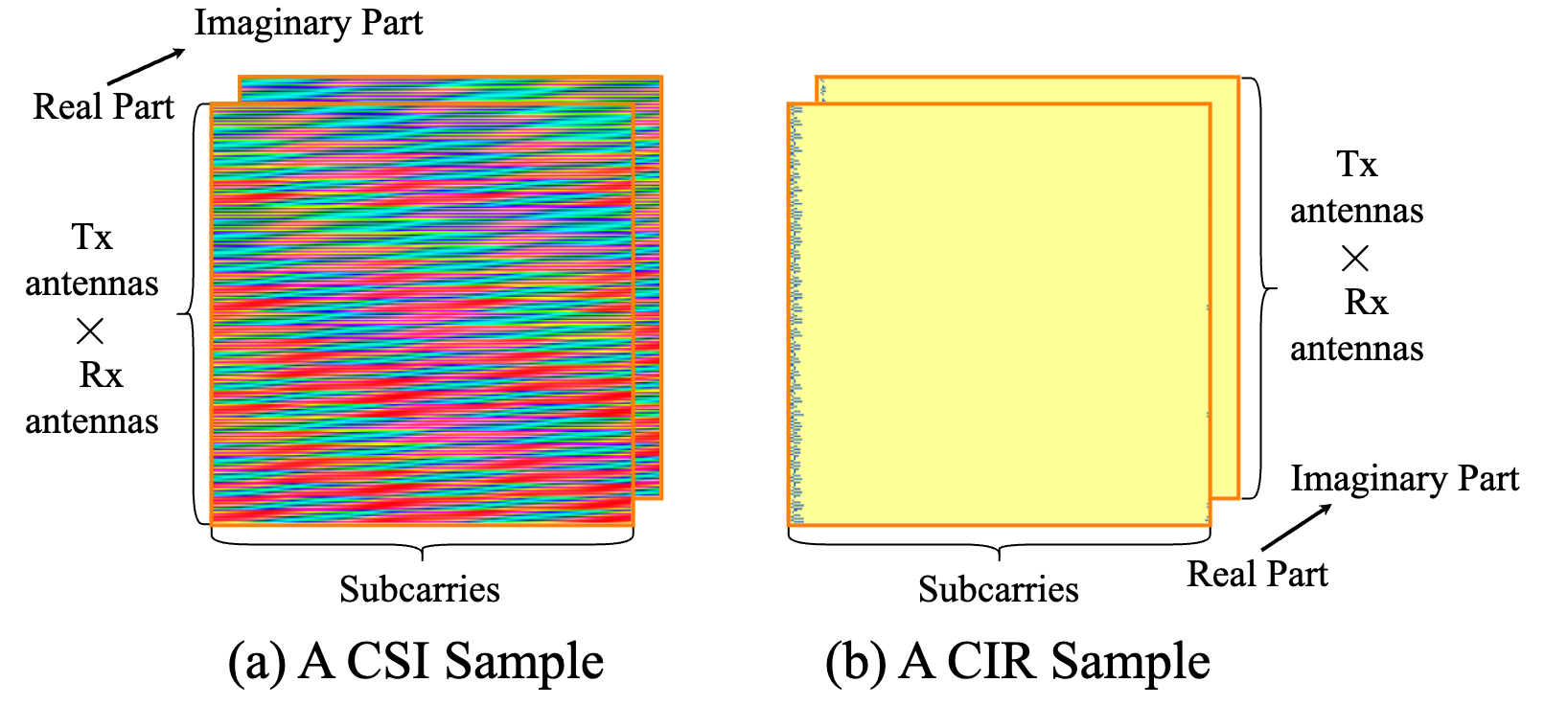}
\caption{Illustration of CSI-CIR correspondence in MIMO channels. Frequency-domain CSI exhibits structured spatial-frequency patterns, while delay-domain CIR reveals sparse multipath components of the same propagation process.}
\label{fig:visual}
\end{figure}

Fig.~\ref{fig:visual} provides an intuitive visualization of the physical motivation behind CSI-CIR contrastive alignment. In MIMO channels, CSI often exhibits structured patterns across antennas and subcarriers due to spatial-frequency correlations, while CIR presents a relatively sparse structure in the delay domain because only a limited number of dominant propagation paths contribute significantly to the received signal. These two representations are not independent modalities, but physically related views of the same propagation process. Therefore, aligning CSI and CIR at the representation level can encourage the model to capture propagation-consistent features, rather than only memorizing local structures in a single domain.

\subsection{Channel Foundation Models}
Building upon SSL, recent studies\cite{palhares2025csi2vec, yang2025wirelessgpt, 11442291} have started to explore CFMs for wireless communications. Different from conventional task-specific neural networks, CFMs aim to learn general-purpose channel representations from large-scale wireless data and adapt them to diverse downstream tasks. This paradigm is particularly suitable for wireless systems, since many tasks, such as channel identification, beam prediction, localization, sensing, channel estimation, CSI feedback, and channel extrapolation, are closely related to the same underlying propagation environment.

Many existing CFMs are built upon reconstruction-based self-supervised objectives, but differ from conventional SSL studies in their emphasis on large-scale pretraining and multi-task transfer. Ott et al.~\cite{ott2024radio} pretrained a Transformer model on 5G channel measurements and finetuned it for indoor localization, showing the potential of pretrained radio representations for reducing the dependence on densely labeled reference points. Large Wireless Model (LWM)~\cite{lwm2024} further explored masked channel modeling on large-scale wireless channel data and adapted the pretrained representations to downstream tasks such as robust beamforming, channel identification, and Sub-6 GHz to mmWave beam prediction. Similar  pretraining ideas have also been extended to CSI feedback, and channel extrapolation tasks, as demonstrated by CSI-MAE~\cite{jiang2026csi}, WiFo~\cite{liu2025wifo}, HeterCSI~\cite{zhang2026hetercsi}, and AirFM-DDA~\cite{bian2026airfm}. These studies show that self-supervised objectives can serve as scalable pretraining tasks for learning transferable channel representations.

In addition to CSI-oriented CFMs, recent studies have extended foundation-model-style pretraining to other wireless signal representations. For example, spectrogram- and IQ-based models, such as LWM-Spectro~\cite{kim2026lwm}, SpectrumFM~\cite{zhou2025spectrumfm}, and IQFM~\cite{mashaal2026iqfm}, learn representations from radio spectrograms or baseband IQ signals and adapt them to tasks including modulation classification, spectrum sensing, anomaly detection, and RF fingerprinting. These works indicate that CFM research is not limited to CSI, but can be generalized to broader radio signal forms for both communication-oriented and sensing-oriented applications.

Beyond single-signal pretraining, another emerging direction is to incorporate discriminative, multimodal, or multitask learning into CFMs. CSI-CLIP~\cite{jiang2025mimo} introduces contrastive alignment between CSI and CIR, showing that physically related channel representations can provide useful self-supervised signals. Multimodal and multitask models, such as WiFo-2~\cite{liu2025foundation}, Wireless World Model~\cite{chen2026wireless}, 6G WavesFM~\cite{aboulfotouh20256g}, and MUSE-FM~\cite{zheng2025muse}, further incorporate heterogeneous information sources, such as CSI, IQ signals, environmental information, point clouds, trajectories, or received symbols. These works suggest that CFMs are evolving from single-domain representation learning toward unified wireless intelligence across communication, sensing, and localization tasks.

Despite these advances, existing CFMs still have several limitations. First, many studies mainly focus on a single-domain signal representation, such as CSI, IQ samples, and spectrograms. However, wireless channels naturally have multiple physically related representations. CSI and CIR describe the same channel from frequency-domain and delay-domain perspectives, respectively, and thus provide complementary information for channel understanding.

Second, many existing pretraining objectives are still dominated by reconstruction-based learning. Although masked reconstruction can effectively exploit local channel correlations, it may also encourage the model to recover missing components through interpolation-like operations, especially in structured MIMO channels. This issue may limit the model's ability to learn high-level and transferable propagation representations.

Third, the physical correspondence between CSI and CIR has not been sufficiently exploited in existing CFMs. Since CSI and CIR are linked by Fourier transformation, their relationship provides a natural self-supervised signal for representation learning. Directly aligning these two domains can encourage the model to learn cross-domain consistent features and improve its adaptability to downstream wireless tasks. This motivates the proposed CSI-CLIP++ framework, which learns transferable channel representations through CSI-CIR contrastive alignment rather than relying only on masked reconstruction.

\begin{figure*}[tbp]
\centering\includegraphics[width=0.9\textwidth]{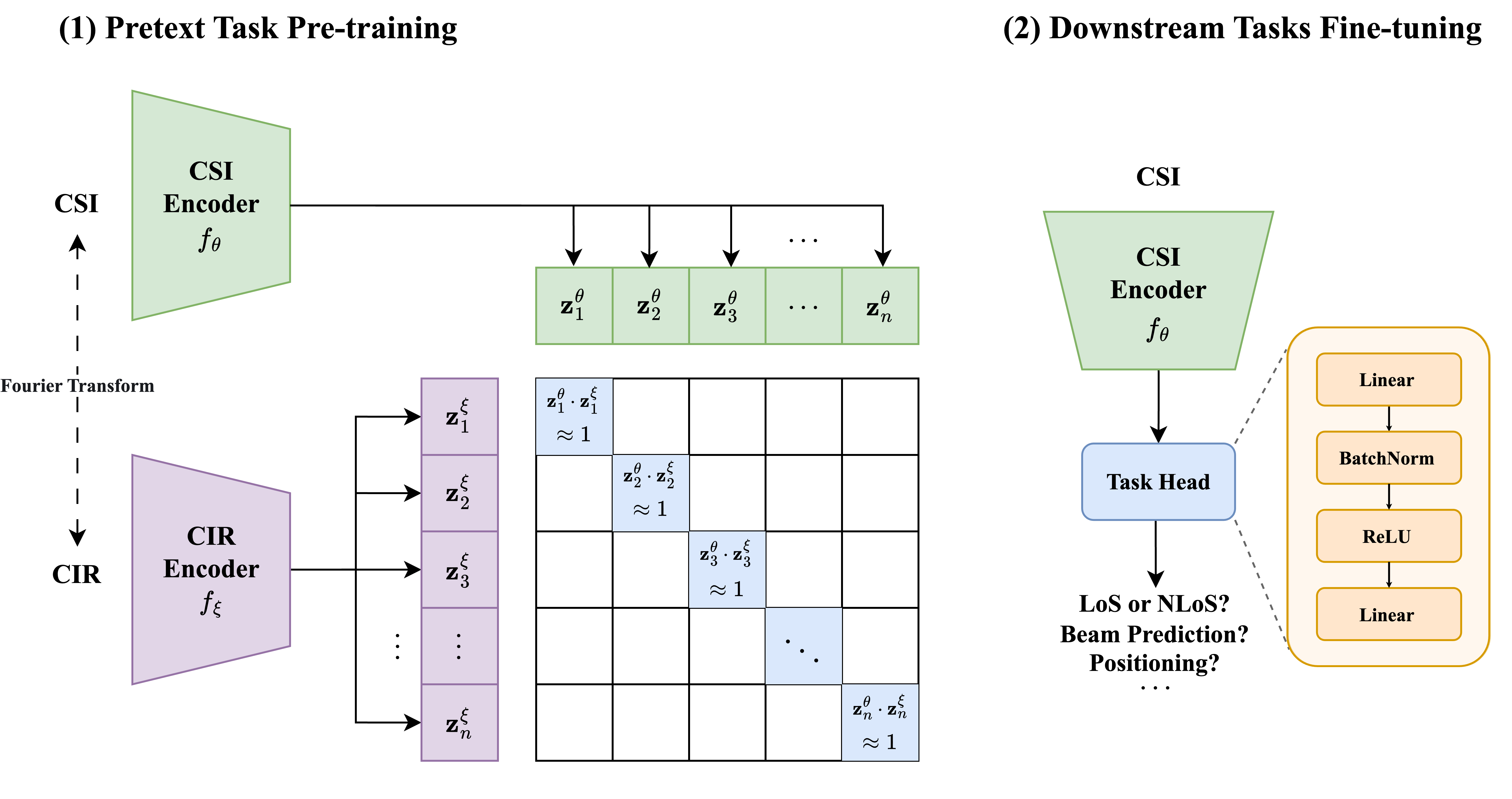}
\caption{Architecture of the proposed CSI-CLIP++.}
\label{fig:pipeline}
\end{figure*}

\section{Task Background}

To evaluate the transferability of the learned channel representations, this paper considers three representative downstream tasks: channel identification, positioning, and beam prediction. These tasks correspond to three typical application directions of CFMs, covering PHY channel understanding, RAN optimization, and ISAC. Specifically, channel identification evaluates whether the learned representation can distinguish different propagation conditions at the PHY level, beam prediction verifies its effectiveness in communication-oriented RAN decision-making, and positioning examines its ability to infer spatial information for ISAC applications. Therefore, these tasks provide a comprehensive evaluation of the generalization ability of the pretrained CSI encoder across different wireless scenarios and prediction types.

\subsection{Channel Identification}
Accurate CSI processing is fundamental to high-performance wireless communications, and channel identification is an important primitive in CSI-based analysis. Channel identification focuses on distinguishing line-of-sight (LoS) and non-line-of-sight (NLoS) propagation conditions\cite{10698216}. This capability is closely related to power control, precoding design, interference mitigation, and reliable resource allocation. However, conventional supervised channel identification models usually rely on labeled channel data collected under specific propagation scenarios and frequency bands. Their performance may degrade when deployed in unseen environments, and the learned knowledge is difficult to reuse for other CSI-related tasks. Therefore, channel identification is used as a representative classification task to evaluate the transferability of the learned channel representation.

Given a CSI sample \(\mathbf{H}_i\), the pretrained CSI encoder first extracts its latent representation, which is then fed into a task-specific classification head. The channel identification task can be formulated as
\begin{equation}
    \hat{y}_i = g_{\phi}^{\mathrm{CI}}(f_{\theta}(\mathbf{H}_i)),
\end{equation}

where \(f_{\theta}\) denotes the pretrained CSI encoder, \(g_{\phi}^{\mathrm{CI}}\) denotes the channel identification head, and \(\hat{y}_i\) is the predicted LoS/NLoS label of the \(i^{\text{th}}\) CSI sample.

\subsection{Beam Prediction}
Beam prediction is essential for high-frequency wireless systems, especially in 5G-Advanced and future 6G networks, where directional beams are required to compensate for severe path loss and support high data rates \cite{bmsurvey}. By selecting an appropriate beam from a predefined codebook, beam prediction helps enhance the received signal strength and reduce beam training overhead. However, large-scale MIMO systems significantly increase the complexity of beam selection, as the number of candidate beams grows with antenna scale and deployment configuration. Existing AI-driven beam prediction methods based on supervised learning usually require separate model design and training for different array scales, frequency bands, and propagation scenarios. This leads to high training costs and limited transferability in dynamic time-varying channel environments \cite{sun2023ai}. Therefore, beam prediction is used to evaluate whether the pretrained CSI encoder can provide effective representations for RAN-oriented tasks.

In this paper, beam prediction is formulated as a multi-class prediction task. Given a CSI sample \(\mathbf{H}_i\), the model predicts the most suitable beam index from a predefined beam codebook:
\begin{equation}
    \hat{b}_i = g_{\phi}^{\mathrm{BP}}(f_{\theta}(\mathbf{H}_i)),
\end{equation}
where \(g_{\phi}^{\mathrm{BP}}\) denotes the beam prediction head, and \(\hat{b}_i\) is the predicted beam index. The predicted beam is selected from the candidate beam set according to the output scores of the task-specific head.

\subsection{Positioning}
Positioning is a core functionality of integrated sensing and communication (ISAC) systems and is expected to play an important role in future 6G networks. High-precision positioning supports indoor navigation, location-based services, and network optimization by providing spatial information about users and environments \cite{isacsurvey}. In addition, sensing capabilities such as positioning, imaging, and environmental reconstruction can further enhance communication functions, including robust beamforming and fast beam failure recovery. Nevertheless, supervised learning-based positioning methods often require large amounts of labeled measurement data across different scenarios and hardware configurations. Their robustness may also be limited in complex NLoS and dense occlusion environments, making generalization across diverse ISAC scenarios challenging.

In this paper, positioning is formulated as a regression task based on CSI representations. Given a CSI sample \(\mathbf{H}_i\), the model predicts the corresponding user location:
\begin{equation}
    \hat{\mathbf{p}}_i = g_{\phi}^{\mathrm{P}}(f_{\theta}(\mathbf{H}_i)),
\end{equation}

where \(g_{\phi}^{\mathrm{P}}\) denotes the positioning head, and \(\hat{\mathbf{p}}_i\) represents the predicted position of the user, \(\hat{\mathbf{p}}_i\) is a two-dimensional coordinate vector.

\section{Proposed framework}

Inspired by contrastive learning frameworks such as CLIP\cite{radford2021clip} and TF-C\cite{tf-c}, this paper proposes CSI-CLIP++, a CSI-CIR contrastive pretraining framework tailored for MIMO wireless channels. Different from general time-frequency contrastive learning, CSI-CLIP++ exploits the physical correspondence between frequency-domain CSI and delay-domain CIR, which are two complementary representations of the same wireless propagation process. By aligning these two domains in a shared embedding space, CSI-CLIP++ aims to learn propagation-consistent channel representations from large-scale unlabeled channel data.

The overall framework of CSI-CLIP++ is illustrated in Fig.~\ref{fig:pipeline}. Given a MIMO channel sample, the frequency-domain CSI and its corresponding delay-domain CIR are first constructed as paired inputs. Two encoders are then used to map CSI and CIR into a common latent space, where paired CSI-CIR samples are pulled closer while unpaired samples are pushed apart through a contrastive objective. After pretraining, the CSI encoder can be adapted to different downstream wireless tasks, including channel identification, beam prediction, and positioning, with lightweight task-specific heads.

\subsection{Pretext Task}
The pretext task of CSI-CLIP++ is formulated as CSI-CIR contrastive alignment. For each channel sample, the frequency-domain CSI and the corresponding delay-domain CIR are treated as a positive pair, since they describe the same wireless propagation process from two different domains. CSI mainly characterizes the spatial-frequency response across antennas and subcarriers, while CIR reveals the delay-domain multipath structure. By learning to associate these two physically related views, the model is encouraged to capture propagation-consistent channel features from unlabeled data.

CSI-CLIP++ adopts a dual-branch encoder structure. The CSI branch and the CIR branch encode their respective inputs independently and project the extracted features into a shared embedding space. The two encoders adopt the same backbone architecture but use separate parameters, allowing each branch to preserve domain-specific characteristics while producing comparable representations for contrastive alignment.

The feature embedding process is formulated as follows:
\begin{equation}
\begin{split}
    \mathbf{z}^{\theta}_{i} &= f_\theta(\mathrm{CSI}_i),\\
    \mathbf{z}^{\xi}_{i} &= f_\xi(\mathrm{CIR}_i),
\end{split}
\end{equation}
where \(f_\theta\) and \(f_\xi\) denote the CSI encoder and the CIR encoder, respectively. \(\mathbf{z}^{\theta}_i\) and \(\mathbf{z}^{\xi}_i\) represent the feature embeddings of the \(i^{\text{th}}\) sample in the shared embedding space.

To align the two channel representations, a contrastive learning objective is introduced. The CSI and CIR embeddings from the same channel sample are regarded as a positive pair, while embeddings from different channel samples are regarded as negative pairs. The objective encourages positive CSI-CIR pairs to have higher similarity than negative pairs, so that the learned representation can preserve the intrinsic relationship between the frequency-domain and delay-domain channel views.

Specifically, cosine similarity is used to measure the similarity between feature embeddings, and a fixed temperature parameter \(\tau\) is used to control the sharpness of the similarity distribution. The contrastive loss is defined as:
\begin{equation}
\mathcal{L} = -\frac{1}{N} \sum_{i=1}^{N} \log 
\frac{\exp(\cos(\mathbf{z}_i^\theta, \mathbf{z}_i^\xi) / \tau)}
{\sum_{j=1}^{N} \exp(\cos(\mathbf{z}_i^\theta, \mathbf{z}_j^\xi) / \tau)},
\label{equ:loss}
\end{equation}
where \(N\) denotes the batch size, and \(\cos(\cdot,\cdot)\) denotes the cosine similarity between two feature vectors. By optimizing this objective, the model learns to associate the CSI representation with its corresponding CIR representation while distinguishing it from non-matching CIR representations in the same batch. In this way, CSI-CLIP++ learns channel embeddings that are consistent across the frequency and delay domains, providing a transferable representation for downstream wireless tasks.

\subsection{Downstream Tasks}
After the CSI-CIR contrastive pretraining stage, the pretrained CSI encoder \(f_\theta\) is transferred to downstream wireless tasks. Since many practical wireless applications rely primarily on CSI as input, the CIR branch is used only during pretraining, while the CSI encoder is retained as the main feature extractor for downstream adaptation.

For each downstream task, a lightweight task-specific head is appended to the output of the pretrained CSI encoder. In this paper, the task-specific head consists of two fully connected layers with a ReLU activation function between them. The resulting model is then fine-tuned using task-specific labeled data. This design allows the downstream model to benefit from the channel representations learned during CSI-CIR pretraining, while adapting the output layer to the specific prediction target of each task.

For channel identification, the task-specific head outputs the predicted category. For beam prediction, the head predicts the appropriate beam index based on the CSI representation. For positioning, the head regresses the user location from the learned channel features. Through these downstream adaptations, the effectiveness and transferability of the pretrained CSI encoder can be evaluated across different wireless communication and sensing tasks.

\section{Experiments}

\subsection{Experimental Setup}

\subsubsection{Dataset}
The experiments are conducted on a large-scale CSI dataset generated from the DeepMIMO platform~\cite{deepmimo}. The pretraining dataset contains more than 0.7 million CSI samples collected from 35 representative wireless communication scenarios. These scenarios cover diverse indoor and outdoor propagation environments, including indoor office, urban cellular, high-frequency communication, and aerial communication settings. The carrier frequencies range from sub-6 GHz to mmWave and terahertz bands, providing diverse channel characteristics for representation learning.

During pretraining, CSI samples from all 35 scenarios are used to learn generalizable channel representations. To reduce the imbalance among scenarios, a stratified sampling strategy is adopted. Specifically, scenarios with relatively small numbers of users are fully included, while subsets are sampled from larger scenarios. This strategy prevents large-scale scenarios from dominating the pretraining process and improves the diversity of the learned channel representations.

To evaluate the transferability of the pretrained CSI encoder, seven unseen representative scenarios are selected for downstream task verification, as shown in Table~\ref{tab:downstream_scenarios}. These scenarios cover both indoor and outdoor environments, with carrier frequencies including 2.4 GHz, 3.5 GHz, 28 GHz, and 60 GHz. Meanwhile, the number of training samples varies from hundreds to tens of thousands, which enables the evaluation of CSI-CLIP++ under different fine-tuning data scales. 
Therefore, the selected scenarios provide a comprehensive benchmark for assessing the effectiveness of the learned channel representations in PHY, RAN, and ISAC.

\begin{table}[tbp]
\centering
\caption{Detailed information of the downstream evaluation scenarios.}
\label{tab:downstream_scenarios}
\resizebox{\columnwidth}{!}{
\begin{tabular}{lccc}
\toprule
Scenario & Frequency & Training Samples & Validation Samples \\
\midrule
Chicago & 3.5 GHz & 228 & 57 \\
San Diego & 3.5 GHz & 1,753 & 439 \\
Charlotte & 3.5 GHz & 2,881 & 721 \\
I1\_2p4 & 2.4 GHz & 15,693 & 3,924 \\
I2\_28B & 28 GHz & 15,702 & 3,926 \\
O1\_3p5B & 3.5 GHz & 38,712 & 9,679 \\
O1\_60 & 60 GHz & 67,017 & 16,755 \\
\bottomrule
\end{tabular}
}
\end{table}


Based on the seven unseen downstream scenarios, we evaluate CSI-CLIP++ in three representative wireless applications: channel identification, beam prediction, and positioning. These tasks correspond to binary classification, multi-class prediction, and regression, respectively, thereby providing a comprehensive evaluation of the learned CSI representations.

\subsection{Implementation Details}

The CSI samples are perfect ray-tracing channel responses generated by DeepMIMO, rather than pilot-estimated noisy CSI. Each complex frequency-domain CSI tensor contains 256 subcarriers, 64 base-station antenna elements from the $8\times8$ UPA, and 4 user-equipment antenna elements from the $2\times2$ UPA; the real and imaginary components are used as model input channels. The maximum number of propagation paths is set to 20, the system bandwidth is 10 MHz. All CSI tensors are normalized before training. The paired CIR input is obtained from the corresponding CSI by applying an inverse discrete Fourier transform, implemented by IFFT, along the subcarrier dimension, so the CSI and CIR branches receive frequency- and delay-domain views of the same channel realization. Unless otherwise specified, SNR-dependent noise effects are not considered, and the evaluation focuses on representation learning under ideal CSI acquisition.

The supervised baseline uses the same model architecture as the pretrained models but is trained from scratch on each downstream task. CSI-CLIP++ uses a ViT-B backbone~\cite{dosovitskiy2020image} to better capture global dependencies across antennas, subcarriers, and channel dimensions.

For fair comparison with reconstruction-based CFMs, we consider both LWM~\cite{lwm2024} and CSI-MAE~\cite{jiang2026csi}. LWM is included in the task-level comparisons under the same downstream scenarios, CSI input configuration, and evaluation metrics whenever applicable. To further isolate the effect of the pretraining objective, we also implement CSI-MAE under the same configuration as CSI-CLIP++: the same ViT-B encoder, the same unlabeled pretraining dataset, the same downstream heads, and the same fine-tuning protocol. CSI-MAE is pretrained by reconstructing masked CSI patches; after pretraining, its decoder is discarded and only the encoder is transferred to downstream tasks.

All models are trained and evaluated on NVIDIA GeForce RTX 4090 GPUs. AdamW is used as the optimizer. The batch size is set to 128, and the maximum number of training epochs is set to 300. Early stopping is applied based on the validation loss to avoid overfitting. The initial learning rate is set to $8\times10^{-4}$ and decayed by 20\% when the validation loss does not improve for ten consecutive epochs.

\subsection{Applications for PHY}
For the PHY application, channel identification is adopted to evaluate whether the pretrained CSI encoder can preserve fundamental propagation characteristics from CSI measurements. Channel identification aims to distinguish different propagation conditions, such as LoS and NLoS links, and therefore provides a direct assessment of the physical discriminability of the learned CSI representation. Compared with beam prediction and positioning, this task focuses more on basic channel structure recognition. Therefore, we evaluate the PHY application by comparing CSI-CLIP++ with the supervised baseline, so as to verify the effectiveness of CSI-CIR contrastive pretraining for channel identification.

The channel identification performance is evaluated using classification accuracy. A higher accuracy indicates that the model can better distinguish different propagation conditions from CSI representations.

To evaluate the effectiveness of pretraining on the PHY task, we compare CSI-CLIP++ with a supervised baseline using the same backbone architecture. The supervised baseline is trained directly on the downstream channel identification task without CSI-CIR contrastive pretraining, while CSI-CLIP++ first learns general CSI representations through pretraining and is then adapted to the downstream task. Therefore, the performance gap between the two methods reflects the benefit introduced by the proposed pretraining strategy.

\begin{table}[tbp]
\centering
\caption{Effectiveness of CSI-CLIP++ pretraining on the PHY channel identification task.}
\label{tab:phy_pretraining_channel_identification}
\resizebox{\columnwidth}{!}{
\begin{tabular}{lccc}
\toprule
Scenario & Supervised & CSI-CLIP++ & Improvement \\
\midrule
Chicago     & 98.25\% & \textbf{100.00\%} & +1.75\% \\
San Diego   & \textbf{100.00\%} & \textbf{100.00\%} & +0.00\% \\
Charlotte   & 98.20\% & \textbf{98.89\%} & +0.69\% \\
I1\_2p4     & \textbf{100.00\%} & \textbf{100.00\%} & +0.00\% \\
I2\_28B     & \textbf{100.00\%} & \textbf{100.00\%} & +0.00\% \\
O1\_3p5B    & 92.11\% & \textbf{99.16\%} & +7.05\% \\
O1\_60      & 97.80\% & \textbf{99.46\%} & +1.66\% \\
\bottomrule
\end{tabular}}
\end{table}

As shown in Table~\ref{tab:phy_pretraining_channel_identification}, CSI-CLIP++ achieves comparable or better channel identification accuracy than the supervised baseline in all scenarios under the same backbone. The gain is most evident in O1\_3p5B, where the accuracy increases from 92.11\% to 99.16\%, while smaller gains are observed in Chicago, Charlotte, and O1\_60. In San Diego, I1\_2p4, and I2\_28B, both methods reach 100\% accuracy, indicating a performance ceiling for this relatively separable LoS/NLoS task. Overall, the results show that CSI-CIR pretraining improves PHY-level discriminability when the supervised baseline is not already saturated.

\subsection{Applications for RAN}

To evaluate the applicability of CSI-CLIP++ to RAN-oriented beam prediction, we conduct two groups of experiments. The first group compares CSI-CLIP++ with a supervised baseline using the same backbone, aiming to verify the effectiveness of the proposed CSI-CIR contrastive pretraining strategy. The second group compares CSI-CLIP++ with existing beam prediction methods on the same dataset, further evaluating its competitiveness against task-specific baselines. The performance is reported using Top-1 and Top-3 accuracy. Top-1 accuracy measures whether the predicted beam exactly matches the optimal beam, while Top-3 accuracy evaluates whether the optimal beam is included among the three most likely candidates.

\subsubsection{Effectiveness of Pretraining}

\begin{table}[!t]
\centering
\caption{Effectiveness of CSI-CLIP++ pretraining for beam prediction. Accuracy is reported in the form of Top-1/Top-3.}
\label{tab:beam_prediction_accuracy}
\resizebox{\columnwidth}{!}{
\begin{tabular}{lccc}
\toprule
Scenario & Supervised & CSI-CLIP++ & Improvement \\
\midrule
Chicago 
& 82.46/87.72 
& \textbf{89.47/94.74}
& +7.01/+7.02 \\

San Diego 
& 68.77/90.37 
& \textbf{75.85/91.11}
& +7.08/+0.74 \\

Charlotte 
& 78.03/93.81 
& \textbf{80.86/94.45}
& +2.83/+0.64 \\

I1\_2p4 
& 95.02/99.38 
& \textbf{95.36/99.64}
& +0.34/+0.26 \\

I2\_28B 
& 79.75/94.16 
& \textbf{83.32/96.75}
& +3.57/+2.59 \\

O1\_3p5B 
& 92.91/98.90 
& \textbf{93.18/99.34}
& +0.27/+0.44 \\

O1\_60 
& 75.29/93.11 
& \textbf{94.60/99.45}
& +19.31/+6.34 \\
\bottomrule
\end{tabular}
}
\end{table}

Table~\ref{tab:beam_prediction_accuracy} compares CSI-CLIP++ with the supervised baseline under the same backbone. CSI-CLIP++ improves both Top-1 and Top-3 accuracy in all seven scenarios, with Top-1 gains ranging from 0.27 to 19.31 percentage points. The largest gain appears in O1\_60, where Top-1 accuracy increases from 75.29\% to 94.60\%, showing that the pretrained representation is especially useful for challenging beam selection cases.

The gains are smaller in I1\_2p4 and O1\_3p5B because the supervised baseline already achieves high Top-3 accuracy. Nevertheless, CSI-CLIP++ still improves Top-1 accuracy, indicating better ranking of the optimal beam rather than only better candidate inclusion. The results also show benefits in both low-sample scenarios, such as Chicago and San Diego, and large-sample scenarios, such as O1\_60, suggesting that the improvement is affected by scenario complexity and beam ambiguity in addition to data scale.

\subsubsection{Comparison with Existing Methods}

\begin{figure*}[!t]
\centering
\includegraphics[width=0.9\textwidth]{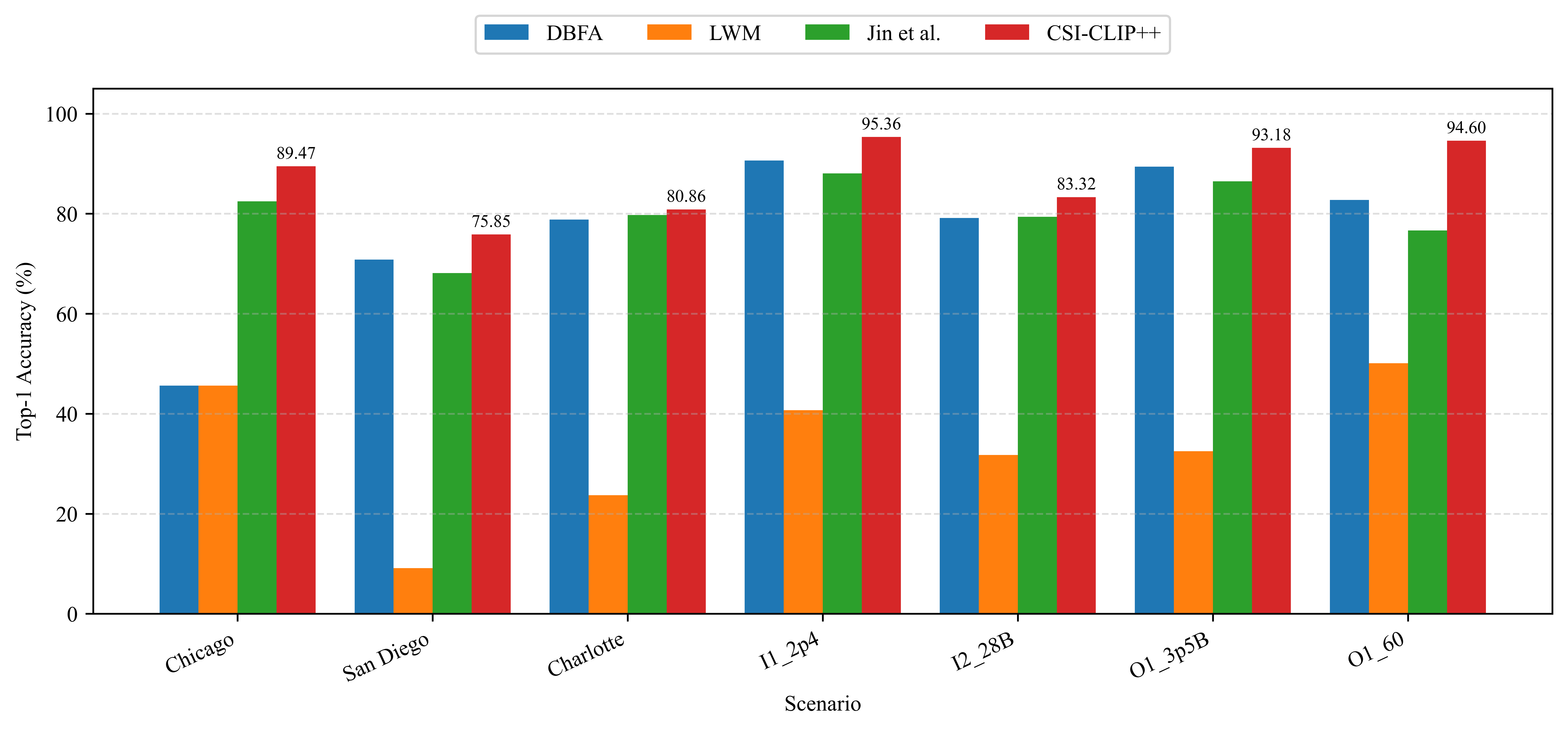}
\caption{Top-1 beam prediction accuracy comparison with existing methods across multiple scenarios.}
\label{fig:beam_existing_methods}
\end{figure*}

Fig.~\ref{fig:beam_existing_methods} compares CSI-CLIP++ with existing beam prediction methods on the same dataset and downstream scenarios. Following the reported results of existing methods, this comparison focuses on Top-1 accuracy under the same beam prediction metric. The compared baselines include DBFA\cite{alrabeiah2020deep}, LWM\cite{lwm2024}, and the method proposed by Jin et al.~\cite{jin2026generalizable}. It is worth noting that the method of Jin et al. incorporates coordinate information for beam prediction, which provides additional spatial prior and usually leads to stronger prediction performance.

CSI-CLIP++ achieves the highest Top-1 accuracy across all seven scenarios, with improvements ranging from 1.11 to 11.87 percentage points over the strongest existing baseline. Although the coordinate-assisted method of Jin et al. obtains competitive results in several scenarios, CSI-CLIP++ still consistently outperforms it and other baselines. This result demonstrates that the proposed CSI-CIR contrastive pretraining can learn highly discriminative and transferable channel representations, enabling strong beam prediction performance without relying on additional coordinate information.

Among all scenarios, the largest gain is observed in O1\_60, where CSI-CLIP++ improves the Top-1 accuracy by 11.87\% over the best existing baseline. This result suggests that CSI-CLIP++ is particularly effective in more challenging scenarios where beam prediction may be affected by complex propagation characteristics or more ambiguous beam selection patterns. Overall, the consistent improvements across different scenarios demonstrate the effectiveness and generalization capability of CSI-CLIP++ for RAN-oriented beam prediction.

\subsection{Applications for ISAC}

\begin{figure*}[tbp]
\centering
\includegraphics[width=\textwidth]{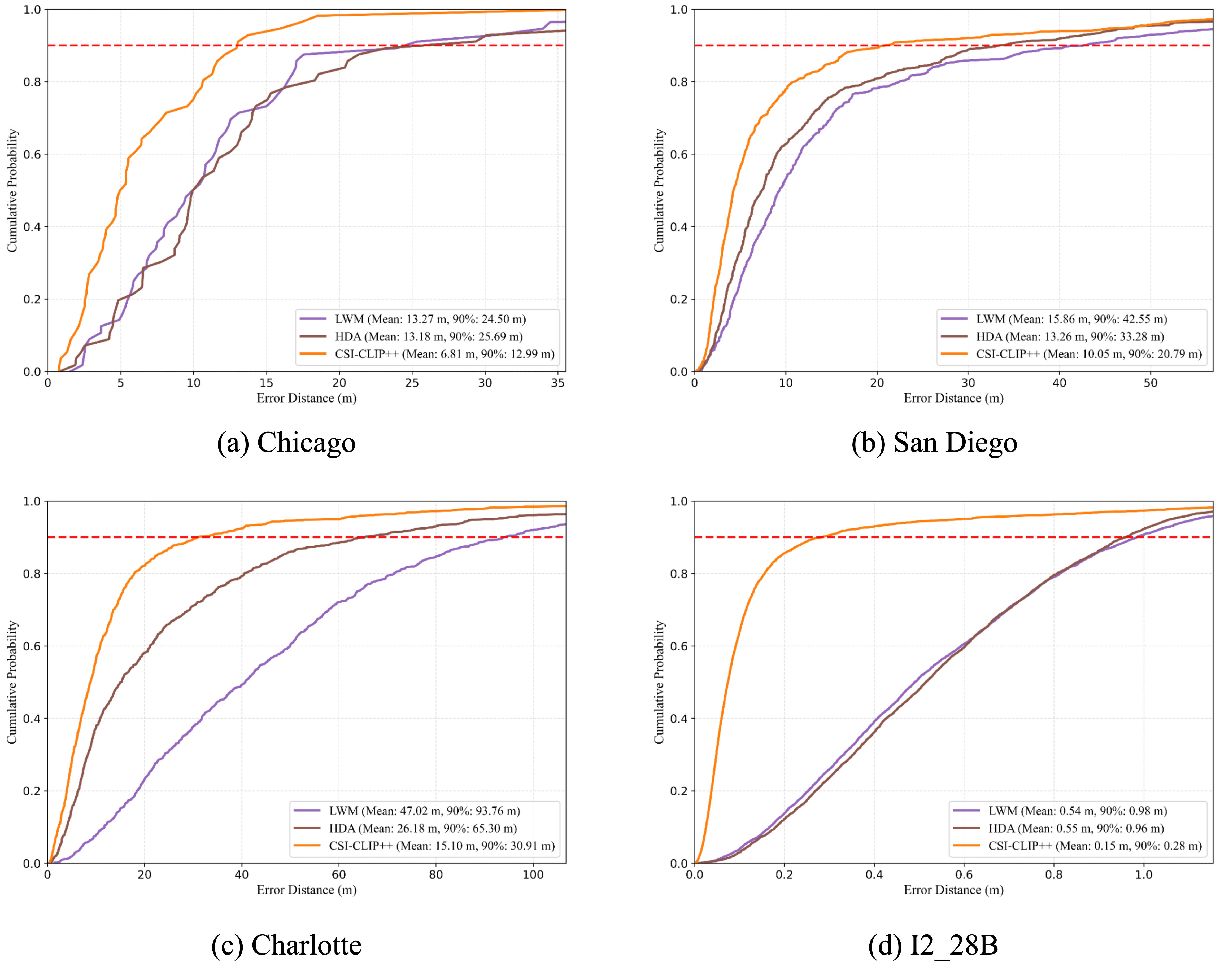}
\caption{CDF curves of the Euclidean positioning error on four representative scenarios.}
\label{fig:isac_positioning_cdf}
\end{figure*}

For the ISAC application, user positioning is adopted to evaluate whether the pretrained CSI encoder can support sensing-oriented spatial inference from CSI measurements. We first examine the benefit of pretraining on seven downstream positioning scenarios, then compare CSI-CLIP++ with existing methods, and finally evaluate its generalization ability on Sionna-based positioning data.

The positioning performance is evaluated by the Euclidean distance between the predicted and ground-truth user coordinates, where a lower value indicates better performance.

\subsubsection{Effectiveness of Pretraining}
To evaluate the effectiveness of pretraining, we compare CSI-CLIP++ with the supervised baseline on seven downstream positioning scenarios. The error reduction is calculated relative to the supervised baseline.

\begin{table}[tbp]
\centering
\caption{Effectiveness of pretraining on the ISAC positioning task evaluated by Euclidean distance.}
\label{tab:isac_pretraining_positioning}
\resizebox{\columnwidth}{!}{
\begin{tabular}{lccc}
\toprule
Scenario & Supervised & CSI-CLIP++ & Error Reduction \\
\midrule
Chicago     & 6.85 m  & \textbf{6.81 m}  & 0.58\% \\
San Diego   & 10.96 m & \textbf{10.05 m} & 8.30\% \\
Charlotte   & 15.99 m & \textbf{15.10 m} & 5.57\% \\
I1\_2p4     & 0.078 m & \textbf{0.076 m} & 2.56\% \\
I2\_28B     & 0.16 m  & \textbf{0.15 m}  & 6.25\% \\
O1\_3p5B    & 3.45 m  & \textbf{2.88 m}  & 16.52\% \\
O1\_60      & 4.25 m  & \textbf{3.63 m}  & 14.59\% \\
\bottomrule
\end{tabular}}
\end{table}

As shown in Table~\ref{tab:isac_pretraining_positioning}, CSI-CLIP++ reduces the positioning error in all seven scenarios. The reductions are small in Chicago and I1\_2p4, where the supervised baseline is already strong, but become more visible in San Diego, O1\_3p5B, and O1\_60, with relative reductions of 8.30\%, 16.52\%, and 14.59\%, respectively. Although the absolute errors differ between indoor and outdoor scenarios due to different spatial ranges, the consistent reductions indicate that CSI-CIR pretraining improves spatial representation learning for ISAC-oriented positioning.

\subsubsection{Comparison with Existing Methods}

To further evaluate the competitiveness of CSI-CLIP++ on the ISAC positioning task, we compare it with two representative methods, LWM\cite{lwm2024} and HDA\cite{chu2024exploiting}, under the same downstream configuration, CSI input setting, and Euclidean-error metric. To provide a more detailed comparison beyond the average error, we plot the cumulative distribution function (CDF) curves of the Euclidean positioning error on four representative scenarios, as shown in Fig.~\ref{fig:isac_positioning_cdf}. A left-shifted CDF curve indicates better positioning performance, since a larger proportion of users can be localized within a smaller error threshold.

As shown in Fig.~\ref{fig:isac_positioning_cdf}, CSI-CLIP++ consistently outperforms LWM and HDA across the four representative scenarios. Its CDF curves are generally shifted to the left, indicating that more users can be localized within smaller Euclidean distance thresholds. This advantage is especially evident in Charlotte and I2\_28B, where CSI-CLIP++ substantially reduces both average errors and large-error cases compared with the competing methods. These results suggest that the pretrained CSI representations provide more effective spatial information for ISAC-oriented positioning.

\subsubsection{Cross-simulator Generalization}

To further examine the generalization ability of the learned CSI representation, we evaluate CSI-CLIP++ on an additional positioning dataset generated by Sionna RT~\cite{Sionna}. The model is pretrained on DeepMIMO and then transferred to the Sionna-generated scenario. This setting introduces a clear domain shift from the DeepMIMO-based pretraining data, including different ray-tracing engines, scene layouts, and channel generation procedures. Therefore, it provides a useful testbed for evaluating whether the pretrained CSI encoder can generalize beyond the original simulator.

\begin{figure}[tbp]
\centering
\includegraphics[width=\columnwidth]{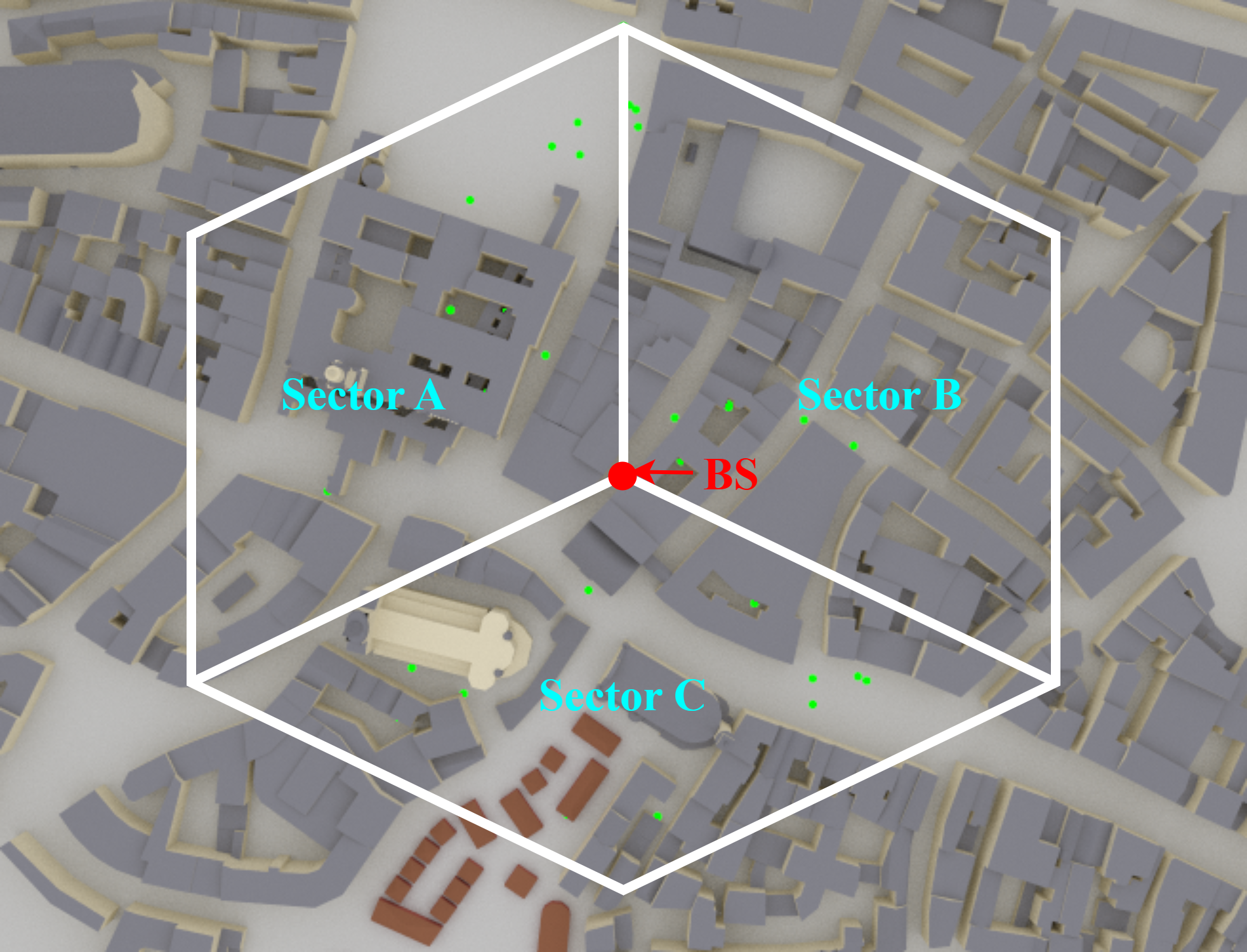}
\caption{Illustration of the Sionna RT etoile urban cellular scenario with three directional sectors.}
\label{fig:sionna_etoile_layout}
\end{figure}

As illustrated in Fig.~\ref{fig:sionna_etoile_layout}, the Sionna RT dataset is generated from an etoile urban cellular scenario with a single-site, three-sector layout. User equipment nodes are sampled at intervals of 1~m, with a maximum base-station-to-user distance of 200~m. The carrier frequency is set to 3.5~GHz, and the maximum number of ray interactions is set to 4. The etoile scenario consists of one base station with three directional sectors, denoted as Sector A, Sector B, and Sector C. Each sector corresponds to a different antenna pointing direction and covers a different user region, leading to different propagation and positioning difficulties.

\begin{table}[tbp]
\centering
\caption{Cross-simulator generalization performance on the Sionna RT positioning dataset evaluated by Euclidean distance.}
\label{tab:sionna_positioning}
\resizebox{\columnwidth}{!}{
\begin{tabular}{lccc}
\toprule
Sector & Supervised & CSI-CLIP++ & Error Reduction \\
\midrule
A & 4.59 m  & \textbf{3.33 m}  & 27.45\% \\
B & 2.75 m  & \textbf{1.67 m}  & 39.27\% \\
C & 11.07 m & \textbf{9.13 m}  & 17.52\% \\
\bottomrule
\end{tabular}}
\end{table}

\begin{table}[tbp]
\centering
\caption{Comparison of different pretraining objectives averaged over seven downstream scenarios.}
\label{tab:pretraining_objective_comparison}
\resizebox{\columnwidth}{!}{
\begin{tabular}{lccc}
\toprule
Method & PHY Acc. $\uparrow$ & RAN Acc. $\uparrow$ & ISAC Error $\downarrow$ \\
\midrule
Supervised & 98.05\% & 81.75\% & 5.96 m \\
CSI-MAE & 99.51\% & 86.84\% & \textbf{5.10 m} \\
CSI-CLIP++ & \textbf{99.64\%} & \textbf{87.52\%} & 5.53 m \\
\bottomrule
\end{tabular}}
\end{table}

\begin{table*}[tbp]
\centering
\caption{Backbone configurations and scaling performance of CSI-CLIP++ on PHY, RAN, and ISAC applications.}
\label{tab:backbone_scaling}
\label{tab:backbone_config}
\resizebox{\textwidth}{!}{
\begin{tabular}{lccccccc}
\toprule
Backbone & Architecture & Depth & Embedding Dim. & Params. & PHY Acc. $\uparrow$ & RAN Acc. $\uparrow$ & ISAC Error $\downarrow$ \\
\midrule
ResNet50 & CNN & 50 layers & 2048 & 23.5M & 98.12\% & 88.24\% & 5.82 m \\
ViT-B/16 & Transformer & 12 layers & 768 & 86.6M & 99.64\% & 87.52\% & 5.53 m \\
ViT-L/16 & Transformer & 24 layers & 1024 & 304.3M & \textbf{99.91\%} & \textbf{89.93\%} & \textbf{4.88 m} \\
\bottomrule
\multicolumn{8}{l}{For ResNet50, the embedding dimension denotes the output feature after global average pooling.}
\end{tabular}}
\end{table*}

As shown in Table~\ref{tab:sionna_positioning}, CSI-CLIP++ reduces the positioning error in all three Sionna RT sectors, although this simulator and layout are not included during DeepMIMO pretraining. The relative reductions are 27.45\%, 39.27\%, and 17.52\% for Sectors A, B, and C, respectively. Sector C remains more difficult in absolute error, but the consistent gains indicate that the learned representation can transfer across ray-tracing engines and layouts under the evaluated setting.

Overall, the ISAC results demonstrate that CSI-CLIP++ can effectively capture spatial channel information for user positioning. Together with the PHY and RAN applications, these results verify the effectiveness of CSI-CLIP++ from multiple perspectives: the PHY application confirms its ability to preserve basic propagation characteristics, the RAN application validates its benefit for communication-oriented beam prediction, and the ISAC application shows its potential for sensing-oriented spatial inference. These observations collectively indicate that CSI-CLIP++ provides a promising foundation-model-style representation learning framework for wireless channel understanding.

\subsection{Comparison with Reconstruction-Based Pretraining}
To distinguish the effect of the pretraining objective from the effect of backbone capacity, we compare supervised training, CSI-MAE, and CSI-CLIP++ under the same ViT-B encoder and the same downstream protocols before the scaling analysis. This experiment is complementary to the following backbone scaling study: here the backbone is fixed and the pretraining objective is changed, whereas the scaling study fixes the CSI-CIR contrastive objective and changes the encoder architecture.
In Table~\ref{tab:pretraining_objective_comparison}, PHY Acc. denotes the average LoS/NLoS channel identification accuracy, RAN Acc. denotes the average Top-1 beam prediction accuracy, and ISAC Error denotes the average Euclidean positioning error over the seven downstream scenarios.

Table~\ref{tab:pretraining_objective_comparison} shows that both self-supervised pretraining objectives improve over direct supervised training on the averaged PHY, RAN, and ISAC metrics. CSI-CLIP++ achieves the best PHY accuracy and RAN beam prediction accuracy, which suggests that CSI-CIR alignment learns more discriminative channel representations for propagation-state recognition and beam selection. This advantage is consistent with the objective design: matching CSI and CIR views encourages the encoder to preserve cross-domain propagation structure rather than only reconstructing local CSI patches.

For ISAC positioning, CSI-MAE achieves a lower average localization error than CSI-CLIP++. This result does not contradict the benefit of CSI-CIR contrastive alignment; rather, it indicates a task-dependent difference between reconstruction and contrastive pretraining. Positioning is a fine-grained regression task that can depend strongly on local spatial-frequency details and absolute CSI variations. The masked reconstruction objective in CSI-MAE may preserve these dense local cues more directly, which can be beneficial for coordinate regression under the evaluated perfect-CSI ray-tracing setting. In contrast, the contrastive objective tends to emphasize representation-level consistency between CSI and CIR, which is more beneficial for discriminative tasks. Therefore, the two objectives show complementary strengths across downstream applications.

\subsection{Scaling Analysis}
To further investigate whether the effectiveness of CSI-CLIP++ is dependent on a specific network architecture, we conduct a backbone scaling experiment across different model structures and capacities. Specifically, ResNet50\cite{resnet}, ViT-B\cite{dosovitskiy2020image}, and ViT-L are selected as representative backbones. ResNet50 represents a convolutional neural network with strong local feature extraction capability, while ViT-B and ViT-L represent Transformer-based architectures with different model capacities. This design enables us to examine both architecture-level variation and capacity-level scaling.

For a fair comparison, all backbones are pretrained using the same CSI-CIR contrastive alignment strategy and then adapted to the same downstream PHY, RAN, and ISAC tasks. The downstream evaluation protocols are kept consistent with the previous experiments. The backbone configurations and the corresponding downstream performance are jointly summarized in Table~\ref{tab:backbone_scaling}.

As shown in Table~\ref{tab:backbone_scaling}, CSI-CLIP++ remains effective across CNN and Transformer backbones. ViT-L achieves the best overall performance, improving PHY accuracy to 99.91\%, RAN accuracy to 89.93\%, and ISAC error to 4.88~m. ResNet50 slightly outperforms ViT-B on the RAN task, suggesting that local convolutional patterns remain useful for beam-related CSI features. Nevertheless, increasing the Transformer capacity from ViT-B to ViT-L improves all three averaged metrics, supporting the scalability of the CSI-CIR contrastive objective across encoder architectures and model capacities.

\section{Conclusions and Future Work}

This paper proposed CSI-CLIP++, a scalable CFM for MIMO wireless channels based on CSI-CIR contrastive alignment. Instead of relying on task-specific labels or reconstructing masked channel components, CSI-CLIP++ exploits the physical correspondence between frequency-domain CSI and delay-domain CIR. By aligning these two complementary channel representations in a shared embedding space, the proposed framework learns cross-domain consistent and task-agnostic channel features from large-scale unlabeled data.

The effectiveness of CSI-CLIP++ was validated on three representative downstream applications. Experimental results show that the pretrained CSI encoder consistently improves downstream performance compared with supervised baselines using the same backbone. In the PHY application, CSI-CLIP++ preserves essential propagation characteristics and achieves comparable or better LoS/NLoS identification accuracy across all evaluated scenarios. In the RAN application, CSI-CLIP++ significantly improves beam prediction accuracy, especially in challenging scenarios where direct supervised training is insufficient. In the ISAC application, CSI-CLIP++ reduces positioning errors across both indoor and outdoor scenarios and also demonstrates clear cross-simulator generalization on the Sionna RT dataset. These results indicate that CSI-CIR contrastive pretraining can provide transferable channel representations for different wireless tasks and deployment conditions. Moreover, the backbone scaling analysis confirms that the proposed framework is not restricted to a specific encoder architecture. Larger Transformer-based backbones achieve the best, suggesting the scalability of CSI-CLIP++ toward larger channel foundation models.

Despite these promising results, several limitations remain to be addressed in future work. First, the current framework is mainly evaluated under fixed antenna and subcarrier configurations. Improving its adaptability to heterogeneous array sizes, bandwidths, and channel dimensions is important for practical deployment. Second, although CSI-CLIP++ shows strong performance after fine-tuning, its frozen or linear-probe transfer ability still needs to be further strengthened, so that the pretrained encoder can be more efficiently adapted to downstream tasks with minimal trainable parameters. Third, the current experiments are mainly based on ray-tracing and simulated CSI data. Future work should further validate the proposed framework on real-world measured channel datasets and investigate the impact of hardware impairments, calibration errors, and environmental dynamics. Finally, extending CSI-CLIP++ to multimodal wireless foundation models by incorporating additional sensing, environmental, or mobility information is a promising direction for building more general AI-native wireless intelligence.

\bibliographystyle{IEEEtran}
\bibliography{bibfile}

\begin{thebibliography}{10}
\providecommand{\url}[1]{#1}
\csname url@samestyle\endcsname
\providecommand{\newblock}{\relax}
\providecommand{\bibinfo}[2]{#2}
\providecommand{\BIBentrySTDinterwordspacing}{\spaceskip=0pt\relax}
\providecommand{\BIBentryALTinterwordstretchfactor}{4}
\providecommand{\BIBentryALTinterwordspacing}{\spaceskip=\fontdimen2\font plus
\BIBentryALTinterwordstretchfactor\fontdimen3\font minus
  \fontdimen4\font\relax}
\providecommand{\BIBforeignlanguage}[2]{{%
\expandafter\ifx\csname l@#1\endcsname\relax
\typeout{** WARNING: IEEEtran.bst: No hyphenation pattern has been}%
\typeout{** loaded for the language `#1'. Using the pattern for}%
\typeout{** the default language instead.}%
\else
\language=\csname l@#1\endcsname
\fi
#2}}
\providecommand{\BIBdecl}{\relax}
\BIBdecl

\bibitem{jiang2025mimo}
J.~Jiang, W.~Yu, Y.~Li, Y.~Gao, and S.~Xu, ``A {MIMO} wireless channel
  foundation model via {CIR}-{CSI} consistency,'' in \emph{2025 IEEE
  International Conference on Machine Learning for Communication and Networking
  (ICMLCN)}.\hskip 1em plus 0.5em minus 0.4em\relax IEEE, May 2025, pp. 1--6.

\bibitem{gao2026ai}
Y.~Gao, Z.~Lu, X.~Wu, W.~Yu, S.~Liu, J.~Du, Y.~Jin, S.~Zhang, X.~Chu, and
  S.~Xu, ``{AI}-driven channel state information ({CSI}) extrapolation for
  {6G}: Current situations, challenges, and future research,'' \emph{IEEE
  Communications Surveys \& Tutorials}, vol.~28, pp. 4485--4518, 2026.

\bibitem{li2025survey}
B.~Li, Q.~Zheng, X.~Tian, M.~Yang, G.~Gui, W.~Jiang, H.~Lei, J.~Jiang, F.~Shu,
  A.~Elhanashi \emph{et~al.}, ``A survey of artificial intelligence enabled
  channel estimation methods: Recent advance, performance, and outlook,''
  \emph{Artificial Intelligence Review}, vol.~58, no.~6, p. 187, Jun. 2025.

\bibitem{pan2025ai}
G.~Pan, Y.~Gao, Y.~Gao, W.~Yu, Z.~Zhong, X.~Yang, X.~Guo, and S.~Xu,
  ``{AI}-driven wireless positioning: Fundamentals, standards,
  state-of-the-art, and challenges,'' \emph{IEEE Communications Surveys \&
  Tutorials}, vol.~28, pp. 4394--4428, 2026.

\bibitem{3gpp38843}
{3GPP}, ``{Study on Artificial Intelligence (AI)/Machine Learning (ML) for NR
  Air Interface},'' {3rd Generation Partnership Project (3GPP)}, Technical
  Report {TR 38.843}, Dec. 2023.

\bibitem{jiang2025towards}
J.~Jiang, Y.~Gao, X.~Wu, and S.~Xu, ``Towards channel foundation models
  ({CFMs}): Motivations, methodologies and opportunities,'' \emph{arXiv
  preprint arXiv:2507.13637}, Jul. 2025.

\bibitem{xu20266g}
S.~Xu and J.~Jiang, ``{6G} native {AI} and channel foundation models,''
  \emph{ZTE Technology Journal}, vol.~32, no.~1, pp. 46--52, Jan. 2026.

\bibitem{radford2021clip}
A.~Radford, J.~W. Kim, C.~Hallacy, A.~Ramesh, G.~Goh, S.~Agarwal, G.~Sastry,
  A.~Askell, P.~Mishkin, J.~Clark, G.~Krueger, and I.~Sutskever, ``Learning
  transferable visual models from natural language supervision,'' in
  \emph{Proceedings of the 38th International Conference on Machine Learning},
  ser. Proceedings of Machine Learning Research, vol. 139, Jul. 2021, pp.
  8748--8763.

\bibitem{xu2025enhanced}
S.~Xu, J.~Jiang, W.~Yu, Y.~Gao, G.~Pan, S.~Mu, Z.~Ai, Y.~Gao, P.~Jiang, and
  C.-X. Wang, ``Enhanced fingerprint-based positioning with practical
  imperfections: Deep learning-based approaches,'' \emph{IEEE Wireless
  Communications}, vol.~33, no.~1, pp. 252--258, Feb. 2026.

\bibitem{shi2025channelmamba}
H.~Shi, K.~Jin, X.~Ren, W.~Li, and Y.~Zhou, ``{ChannelMamba}: A {Mamba}-driven
  selective state-space model for channel prediction of high-mobility {MIMO} in
  {6G} {IoT},'' \emph{IEEE Transactions on Wireless Communications}, vol.~25,
  pp. 5291--5305, 2026.

\bibitem{wen2018deep}
C.-K. Wen, W.-T. Shih, and S.~Jin, ``Deep learning for massive {MIMO} {CSI}
  feedback,'' \emph{IEEE Wireless Communications Letters}, vol.~7, no.~5, pp.
  748--751, Oct. 2018.

\bibitem{jiang2025mtca}
J.~Jiang, W.~Yu, Y.~Gao, and S.~Xu, ``{MTCA}: Multi-task channel analysis for
  wireless communication,'' in \emph{2025 IEEE 102nd Vehicular Technology
  Conference (VTC2025-Fall)}.\hskip 1em plus 0.5em minus 0.4em\relax IEEE,
  Sept. 2025, pp. 1--6.

\bibitem{wang2023super}
X.~Wang, K.~Guan, D.~He, Z.~Zhang, H.~Zhang, J.~Dou, and Z.~Zhong,
  ``Super-resolution of wireless channel characteristics: A multitask learning
  model,'' \emph{IEEE Transactions on Antennas and Propagation}, vol.~71,
  no.~10, pp. 8197--8209, Oct. 2023.

\bibitem{jagannath2021multi}
A.~Jagannath and J.~Jagannath, ``Multi-task learning approach for automatic
  modulation and wireless signal classification,'' in \emph{ICC 2021 - IEEE
  International Conference on Communications}.\hskip 1em plus 0.5em minus
  0.4em\relax IEEE, Jun. 2021, pp. 1--7.

\bibitem{zhu2026fm}
G.~Zhu, Y.~Hu, S.~Jayaweera, W.~Gao, W.-H. Wang, J.~Zhang, B.~Wang, C.~Wu, and
  K.~J. Liu, ``{AM}-{FM}: A foundation model for ambient intelligence through
  {WiFi},'' \emph{arXiv preprint arXiv:2602.11200}, Feb. 2026.

\bibitem{pan2025large}
G.~Pan, K.~Huang, H.~Chen, S.~Zhang, C.~H{\"a}ger, and H.~Wymeersch, ``Large
  wireless localization model ({LWLM}): A foundation model for positioning in
  {6G} networks,'' \emph{arXiv preprint arXiv:2505.10134}, May 2025.

\bibitem{guler2026multi}
B.~Guler, G.~Geraci, and H.~Jafarkhani, ``A multi-task foundation model for
  wireless channel representation using contrastive and masked autoencoder
  learning,'' \emph{IEEE Journal on Selected Areas in Communications}, vol.~44,
  pp. 4489--4504, 2026.

\bibitem{jiao2025addressing}
T.~Jiao, Z.~Xiao, Y.~Xu, C.~Ye, Y.~Huang, Z.~Chen, L.~Cai, J.~Chang, D.~He,
  Y.~Guan \emph{et~al.}, ``Addressing the curse of scenario and task
  generalization in {AI}-{6G}: A multi-modal paradigm,'' \emph{IEEE
  Transactions on Wireless Communications}, vol.~24, no.~9, pp. 7377--7391,
  Sept. 2025.

\bibitem{cheng2025foundation}
X.~Cheng, B.~Liu, X.~Liu, E.~Liu, and Z.~Huang, ``Foundation model empowered
  synesthesia of machines ({SoM}): {AI}-native intelligent multi-modal
  sensing-communication integration,'' \emph{IEEE Transactions on Network
  Science and Engineering}, vol.~13, pp. 762--782, 2026.

\bibitem{gui2024survey}
J.~Gui, T.~Chen, J.~Zhang, Q.~Cao, Z.~Sun, H.~Luo, and D.~Tao, ``A survey on
  self-supervised learning: Algorithms, applications, and future trends,''
  \emph{IEEE Transactions on Pattern Analysis and Machine Intelligence},
  vol.~46, no.~12, pp. 9052--9071, Dec. 2024.

\bibitem{xie2022simmim}
Z.~Xie, Z.~Zhang, Y.~Cao, Y.~Lin, J.~Bao, Z.~Yao, Q.~Dai, and H.~Hu,
  ``{SimMIM}: A simple framework for masked image modeling,'' in
  \emph{Proceedings of the IEEE/CVF Conference on Computer Vision and Pattern
  Recognition}, Jun. 2022, pp. 9653--9663.

\bibitem{wang2023droppos}
H.~Wang, J.~Fan, Y.~Wang, K.~Song, T.~Wang, and Z.-X. Zhang, ``{DropPos}:
  Pre-training vision transformers by reconstructing dropped positions,''
  \emph{Advances in Neural Information Processing Systems}, vol.~36, pp.
  46\,134--46\,151, Dec. 2023.

\bibitem{he2020momentum}
K.~He, H.~Fan, Y.~Wu, S.~Xie, and R.~Girshick, ``Momentum contrast for
  unsupervised visual representation learning,'' in \emph{Proceedings of the
  IEEE/CVF Conference on Computer Vision and Pattern Recognition}, Jun. 2020,
  pp. 9729--9738.

\bibitem{simclr}
T.~Chen, S.~Kornblith, M.~Norouzi, and G.~Hinton, ``A simple framework for
  contrastive learning of visual representations,'' in \emph{Proceedings of the
  37th International Conference on Machine Learning}, ser. Proceedings of
  Machine Learning Research, vol. 119, Jul. 2020, pp. 1597--1607.

\bibitem{assran2023self}
M.~Assran, Q.~Duval, I.~Misra, P.~Bojanowski, P.~Vincent, M.~Rabbat, Y.~LeCun,
  and N.~Ballas, ``Self-supervised learning from images with a joint-embedding
  predictive architecture,'' in \emph{Proceedings of the IEEE/CVF Conference on
  Computer Vision and Pattern Recognition}, Jun. 2023, pp. 15\,619--15\,629.

\bibitem{devlin2019bert}
J.~Devlin, M.-W. Chang, K.~Lee, and K.~Toutanova, ``{BERT}: Pre-training of
  deep bidirectional transformers for language understanding,'' in
  \emph{Proceedings of the 2019 Conference of the North American Chapter of the
  Association for Computational Linguistics: Human Language Technologies},
  vol.~1, Jun. 2019, pp. 4171--4186.

\bibitem{he2022masked}
K.~He, X.~Chen, S.~Xie, Y.~Li, P.~Doll{\'a}r, and R.~Girshick, ``Masked
  autoencoders are scalable vision learners,'' in \emph{Proceedings of the
  IEEE/CVF Conference on Computer Vision and Pattern Recognition}, Jun. 2022,
  pp. 16\,000--16\,009.

\bibitem{hondru2025masked}
V.~Hondru, F.~A. Croitoru, S.~Minaee, R.~T. Ionescu, and N.~Sebe, ``Masked
  image modeling: A survey,'' \emph{International Journal of Computer Vision},
  vol. 133, no.~10, pp. 7154--7200, Oct. 2025.

\bibitem{raviv2026wimamba}
T.~Raviv and N.~Shlezinger, ``{WiMamba}: Linear-scale wireless foundation
  model,'' \emph{arXiv preprint arXiv:2603.26367}, Mar. 2026.

\bibitem{sheng2025wireless}
Y.~Sheng, J.~Wang, X.~Zhou, L.~Liang, H.~Ye, S.~Jin, and G.~Y. Li, ``A wireless
  foundation model for multi-task prediction,'' \emph{arXiv preprint
  arXiv:2507.05938}, Jul. 2025.

\bibitem{catak2025bert4mimo}
F.~O. Catak, M.~Kuzlu, and U.~Cali, ``{BERT4MIMO}: A foundation model using
  {BERT} architecture for massive {MIMO} channel state information
  prediction,'' \emph{arXiv preprint arXiv:2501.01802}, Jan. 2025.

\bibitem{lwm2024}
S.~Alikhani, G.~Charan, and A.~Alkhateeb, ``Lwm: A pre-trained wireless
  foundation model for universal feature extraction,'' in \emph{2025 IEEE
  International Conference on Machine Learning for Communication and Networking
  (ICMLCN)}, 2025, pp. 1--6.

\bibitem{jiang2026csi}
J.~Jiang, X.~Ruan, and S.~Xu, ``{CSI}-{MAE}: A masked autoencoder-based channel
  foundation model,'' \emph{arXiv preprint arXiv:2601.03789}, Jan. 2026.

\bibitem{hu2024comprehensive}
H.~Hu, X.~Wang, Y.~Zhang, Q.~Chen, and Q.~Guan, ``A comprehensive survey on
  contrastive learning,'' \emph{Neurocomputing}, vol. 610, p. 128645, Dec.
  2024.

\bibitem{palhares2025csi2vec}
V.~Palhares, S.~Taner, and C.~Studer, ``{CSI2Vec}: Towards a universal {CSI}
  feature representation for positioning and channel charting,'' \emph{arXiv
  preprint arXiv:2506.05237}, Jun. 2025.

\bibitem{yang2025wirelessgpt}
T.~Yang, P.~Zhang, M.~Zheng, Y.~Shi, L.~Jing, J.~Huang, and N.~Li,
  ``{WirelessGPT}: A generative foundation model for multi-task integrated
  sensing and communication,'' \emph{IEEE Journal on Selected Areas in
  Communications}, vol.~44, pp. 2259--2273, 2026.

\bibitem{11442291}
J.~Guo, Z.~Deng, Z.~Qiao, J.~Zhang, J.~Xue, D.~Niyato, and Z.~Xu, ``Scalable
  pre-trained masked channel model of wireless communications,'' \emph{IEEE
  Transactions on Communications}, vol.~74, pp. 6197--6212, 2026.

\bibitem{ott2024radio}
J.~Ott, J.~Pirkl, M.~Stahlke, T.~Feigl, and C.~Mutschler, ``Radio foundation
  models: Pre-training transformers for {5G}-based indoor localization,'' in
  \emph{2024 International Conference on Indoor Positioning and Indoor
  Navigation (IPIN)}.\hskip 1em plus 0.5em minus 0.4em\relax IEEE, Oct. 2024,
  pp. 1--6.

\bibitem{liu2025wifo}
B.~Liu, S.~Gao, X.~Liu, X.~Cheng, and L.~Yang, ``{WiFo}: Wireless foundation
  model for channel prediction,'' \emph{Science China Information Sciences},
  vol.~68, no.~6, p. 162302, Jun. 2025.

\bibitem{zhang2026hetercsi}
C.~Zhang, X.~Lyu, C.~Ren, S.~Liu, Q.~Cui, and X.~Tao, ``{HeterCSI}:
  Channel-adaptive heterogeneous {CSI} pretraining framework for generalized
  wireless foundation models,'' \emph{arXiv preprint arXiv:2601.18200}, Jan.
  2026.

\bibitem{bian2026airfm}
K.~Bian, M.~Tao, J.~Mo, Z.~Chen, and L.~Chen, ``{AirFM}-{DDA}: Air-interface
  foundation model in the delay-doppler-angle domain for {AI}-native {6G},''
  \emph{arXiv preprint arXiv:2605.00020}, May 2026.

\bibitem{kim2026lwm}
N.~Kim, S.~Alikhani, and A.~Alkhateeb, ``{LWM}-spectro: A foundation model for
  wireless baseband signal spectrograms,'' \emph{arXiv preprint
  arXiv:2601.08780}, Jan. 2026.

\bibitem{zhou2025spectrumfm}
F.~Zhou, C.~Liu, H.~Zhang, W.~Wu, Q.~Wu, T.~Q.~S. Quek, and C.-B. Chae,
  ``{SpectrumFM}: A foundation model for intelligent spectrum management,''
  \emph{IEEE Journal on Selected Areas in Communications}, vol.~44, pp.
  4471--4488, 2026.

\bibitem{mashaal2026iqfm}
O.~Mashaal and H.~Abou-Zeid, ``{IQFM}: A wireless foundation model for {I}/{Q}
  streams in {AI}-native {6G},'' \emph{IEEE Open Journal of the Communications
  Society}, vol.~7, pp. 1426--1441, 2026.

\bibitem{liu2025foundation}
B.~Liu, X.~Liu, S.~Gao, X.~Cheng, and L.~Yang, ``Foundation model for
  intelligent wireless communications,'' \emph{arXiv preprint
  arXiv:2511.22222}, Nov. 2025.

\bibitem{chen2026wireless}
Z.~Chen, Y.~Ren, Y.~Huang, Q.~Sun, N.~Li, Y.~Huang, Y.~Li, L.~Xia
  \emph{et~al.}, ``A wireless world model for {AI}-native {6G} networks,''
  \emph{arXiv preprint arXiv:2603.25216}, Mar. 2026.

\bibitem{aboulfotouh20256g}
A.~Aboulfotouh, E.~Mohammed, and H.~Abou-Zeid, ``{6G} {WavesFM}: A foundation
  model for sensing, communication, and localization,'' \emph{IEEE Open Journal
  of the Communications Society}, vol.~6, pp. 6792--6807, 2025.

\bibitem{zheng2025muse}
T.~Zheng, J.~Guo, L.~Dai, S.~Jin, and J.~Zhang, ``{MUSE}-{FM}: Multi-task
  environment-aware foundation model for wireless communications,'' \emph{arXiv
  preprint arXiv:2509.01967}, Sept. 2025.

\bibitem{10698216}
P.~Agarwal, R.~Kumar, D.~K. Jhariya, and M.~K. Singh, ``Comparative analysis of
  machine learning algorithms for {LOS}/{NLOS} identification,'' in \emph{2024
  First International Conference on Electronics, Communication and Signal
  Processing (ICECSP)}.\hskip 1em plus 0.5em minus 0.4em\relax IEEE, 2024, pp.
  1--5.

\bibitem{bmsurvey}
Q.~Xue, C.~Ji, S.~Ma, J.~Guo, Y.~Xu, Q.~Chen, and W.~Zhang, ``A survey of beam
  management for mmwave and {THz} communications towards {6G},'' \emph{IEEE
  Communications Surveys \& Tutorials}, vol.~26, no.~3, pp. 1520--1559, Third
  Quarter 2024.

\bibitem{sun2023ai}
C.~Sun, L.~Zhao, T.~Cui, H.~Li, Y.~Bai, S.~Wu, and Q.~Tong, ``{AI} model
  selection and monitoring for beam management in {5G}-advanced,'' \emph{IEEE
  Open Journal of the Communications Society}, vol.~5, pp. 38--50, Jan. 2024.

\bibitem{isacsurvey}
A.~Magbool, V.~Kumar, Q.~Wu, M.~Di~Renzo, and M.~F. Flanagan, ``A survey on
  integrated sensing and communication with intelligent metasurfaces: Trends,
  challenges, and opportunities,'' Jan. 2024.

\bibitem{tf-c}
X.~Zhang, Z.~Zhao, T.~Tsiligkaridis, and M.~Zitnik, ``Self-supervised
  contrastive pre-training for time series via time-frequency consistency,''
  \emph{Advances in Neural Information Processing Systems}, vol.~35, pp.
  3988--4003, Dec. 2022.

\bibitem{deepmimo}
A.~Alkhateeb, ``{DeepMIMO}: A generic deep learning dataset for millimeter wave
  and massive {MIMO} applications,'' in \emph{2019 Information Theory and
  Applications Workshop (ITA)}, San Diego, CA, USA, Feb. 2019, pp. 1--8.

\bibitem{dosovitskiy2020image}
A.~Dosovitskiy, L.~Beyer, A.~Kolesnikov, D.~Weissenborn, X.~Zhai,
  T.~Unterthiner, M.~Dehghani, M.~Minderer, G.~Heigold, S.~Gelly \emph{et~al.},
  ``An image is worth {16x16} words: Transformers for image recognition at
  scale,'' \emph{arXiv preprint arXiv:2010.11929}, Oct. 2020.

\bibitem{alrabeiah2020deep}
M.~Alrabeiah and A.~Alkhateeb, ``Deep learning for mmwave beam and blockage
  prediction using sub-{6} {GHz} channels,'' \emph{IEEE Transactions on
  Communications}, vol.~68, no.~9, pp. 5504--5518, Sept. 2020.

\bibitem{jin2026generalizable}
Y.~Jin, Y.~Li, J.~Jiang, Y.~Gao, S.~Liu, J.~Du, Z.~Yang, and S.~Xu,
  ``Generalizable and robust beam prediction for {6G} networks: A deep-learning
  framework with positioning feature fusion,'' \emph{arXiv preprint
  arXiv:2602.09685}, Feb. 2026.

\bibitem{chu2024exploiting}
L.~Chu, A.~Alghafis, and A.~F. Molisch, ``Exploiting semantic localization in
  highly dynamic wireless networks using deep homoscedastic domain
  adaptation,'' \emph{IEEE Transactions on Communications}, vol.~73, no.~3, pp.
  2032--2046, Mar. 2025.

\bibitem{Sionna}
J.~Hoydis, S.~Cammerer, F.~{Ait Aoudia}, A.~Vem, N.~Binder, G.~Marcus, and
  A.~Keller, ``{Sionna}: An open-source library for next-generation physical
  layer research,'' \emph{arXiv preprint arXiv:2203.11854}, Mar. 2022.

\bibitem{resnet}
K.~He, X.~Zhang, S.~Ren, and J.~Sun, ``Deep residual learning for image
  recognition,'' in \emph{Proceedings of the IEEE Conference on Computer Vision
  and Pattern Recognition}, Jun. 2016, pp. 770--778.

\end{thebibliography}
\balance

\end{document}